\documentclass[runningheads]{llncs}
\usepackage{graphicx}
\begin{document}
\title{Temporal Aspects of Smart Contracts for Financial Derivatives}
%
%
\author{Christopher D. Clack
\and
Gabriel Vanca 
}
\authorrunning{C. Clack and G.Vanca}
%
\institute{Centre for Blockchain Technologies, \break 
Department of Computer Science, \break 
University College London \break
\email{clack@cs.ucl.ac.uk}\\
\url{http://www.cs.ucl.ac.uk/staff/C.Clack} 
}
\maketitle              
\begin{abstract}


Implementing smart contracts to automate the performance of high-value over-the-counter (OTC) financial derivatives is a formidable challenge. 
Due to the regulatory framework and the scale of financial risk if a contract were to go wrong, the performance of these contracts
must be enforceable in law and there is an absolute requirement 
that the smart contract
will be faithful to the intentions of the parties as expressed in the original legal documentation.
Formal methods provide an attractive route for validation and assurance, 
and here we present early results from an investigation
of the semantics of industry-standard legal documentation for OTC derivatives.  We explain
the need for a formal representation that combines temporal, deontic and operational aspects,
and focus on the requirements for the {\em temporal} aspects as derived from the legal text.
The relevance of this work extends beyond OTC derivatives and is applicable to understanding the temporal semantics
of a wide range of legal documentation.



\end{abstract}
\section{Introduction}

Current research on smart contracts includes a range of use cases, from straightforward
automation of relatively simple and relatively low-value business processes to the
automation of large and complex legal agreements that have extremely high value
and may last for decades.  The automation of OTC derivatives contracts
lies at the latter end of that range, and substantial research and development in this area
has been occurring within universities, investment banks, law firms and financial services
trade associations for several years.

Here we use the term {\em smart legal contract} to refer to a legal contract whose 
performance is automated on distributed ledger technology, and the term {\em smart contract code}
to refer to the code that automates the legal contract \cite{stark}.  In some related research the 
term {\em smart contract}  refers only to the code, yet that definition is problematic in this context since
the code itself may not be a legal contract.  There are many examples of problems with
terminology when computer scientists, lawyers and banking technologists work together \cite{SCT-R3-3}
and we therefore use
a portmanteau definition of the term {\em smart contract} as follows \cite{SCT2016}:

\begin{quote}
{\em A smart contract is an automatable and
enforceable agreement. Automatable by computer,
although some parts may require human
input and control. Enforceable either by legal
enforcement of rights and obligations or via
tamper-proof execution of computer code.}
\end{quote}

The above definition contains key elements for automating OTC derivatives: 
first, the enforcement of rights and obligations by recourse to a court of law is essential
because of the regulatory framework (and the scale of financial risk involved);
second, the smart contract
code may require human input and control, for example where the code encounters a state that requires
human discretion to decide how to proceed or where it is necessary to pause and modify or cancel
the code due to changes in the law.

There is an absolute requirement 
that the smart contract
will be faithful to the intentions of the parties as expressed in the original legal documentation; hence our interest in
the use of formal methods to validate the smart contract code. There is also a requirement that the processes involved
in automating OTC derivatives (including the production of smart contract code and the validation of that code)
align with current standardised workflow in terms of how the legal documentation is structured and negotiated.  
The context of our work is therefore the Smart Contract Templates project \cite{smartcontracttemplates,SCT2016,SCT2016a} 
which focuses on alignment with standard practice, including greater standardisation of smart contract code.

Our aim is to derive a formal semantic representation of the set of documents that comprise the legal agreement underlying
each individual OTC derivatives transaction.  Each transaction will be automated by a separate process (an instance of the 
smart contract code).  
Our initial examination of the documentation \cite{SCT-R3-3,ClackJDB} has demonstrated the need for 
a combined semantic specification, including at least the temporal, deontic and operational aspects of the legal agreement.
Once this has been achieved, there are two possible routes for validation: either (i) validation scenarios could be generated
from the formal specification and used during verification and validation of the smart contract code;\footnote{For example, ``what if?'' scenarios might 
posit a sequence of actions by the parties,
or possible changes in the law during running of the code, together with the required outcome.} or 
(ii) a semantic specification of the smart contract code could be checked against the semantics of the legal documentation
(perhaps achievable automatically, at least in part).

Here we present 
early results from our investigation of the temporal semantics of the legal documentation for OTC derivatives.  

This paper aims to be accessible not only to academics but also to practitioners such as banking techologists, lawyers, and regulators.
Although we present early results from a study of the semantics of legal documentation for OTC derivatives, our observations have much broader implications for
the use of formal methods in representing the semantics of many types of legal documentation.

%

\subsection{Standardisation of OTC Derivatives Contracts}

OTC derivatives are often purchased 
as a mechanism for risk management so that the precise form of the purchased derivative will match the purchaser's financial exposures.  These derivatives contracts can have substantial value, complexity and longevity,\footnote{Derivatives contracts often last 5 years and can last as long as 30 years.} and a firm that purchases a bespoke derivative will need legal clarity and protection relating to the terms of the agreement.  

Negotiating the terms and conditions of bespoke derivatives contracts can itself be a lengthy and costly process.
This complexity and cost can be improved by increasing standardisation of that process.  The International Swaps and Derivatives Association (ISDA) provides a set of legal templates that are commonly used as a known basis for negotiation between counterparties.  The primary template is the ISDA Master Agreement, which covers a range of derivatives from ``vanilla'' interest rate swaps to complex options contracts and which can be used for multicurrency and cross-border transactions.\footnote{There are two versions in common use: the 1992 ISDA Master Agreement and the 2002 ISDA Master Agreement.}

The Master Agreement contains standard clauses that are generally non-contentious.  However, there is a need for customised clauses to be added, and this is achieved using a Schedule template which sets out those areas that are typically customised (and additional clauses may be added).  A Credit Support Annex might also be added if the bank requires the firm to provide collateral to reduce its credit risk to the bank.

After these documents have been agreed and signed by the counterparties,\footnote{Derivatives contracts may  involve more than two parties.} they constitute a single agreement and the counterparties may enter into one or more derivatives transactions based on that agreement.  Each such transaction is specified using a written or electronic Confirmation document setting out the economic terms for that individual transaction, and that Confirmation document is considered to be part of the overall agreement.\footnote{It may also be possible to attach additional terms and conditions, including additional credit support, to individual transactions.}

\subsection{Smart Contract Templates}

The Smart Contract Templates project \cite{smartcontracttemplates,SCT2016,SCT2016a} addresses the process
of writing, testing and debugging the smart contract code \cite{SCT2016} that will perform 
a complex OTC derivatives contract in an automated fashion on a suitable technology platform, and how to align that process with the process 
of using the ISDA document set.  In this paper such legal agreements are called ``smart OTC derivatives''.

Smart Contract Templates provide standardised smart contract code
``templates'' for the ISDA document set (the Master Agreement, Schedule, Credit Suport Annex).
The name ``template'' indicates that the code will leave some terms as yet undefined (example values may be inserted for the purposes 
of testing and debugging).  These templates may be developed and comprehensively verified and validated in advance, and this might include different versions of the code designed to run on different technology platforms.

The workflow for Smart Contract Templates matches that for the institutional workflow as follows:
\begin{itemize}
\item
A Smart Contract Template will have been developed, tested and debugged in advance for the standard ISDA Master Agreement and Schedule.\footnote{For the rest of this paper we assume the 2002 ISDA document set.}
\item
When two or more counterparties establish an agreement, they may negotiate modifications to the Schedule.  When the negotiation is complete, a copy will be made of the existing Smart Contract Template and the smart contract code in this copy will be modified --- many of the undefined terms in the template will be bound to appropriate values (e.g. the counterparty details), and depending on the extent of the modifications to the legal text, this may require a more or less substantial rewriting of the smart contract code.  The overall structure of the code should remain the same, but the code is likely to require further verification and validation.  This process should benefit from the fact that the previous template had already been verified and validated.

The resulting modified smart contract code should accurately reflect the intentions of the parties 
under the agreement, but it is not yet ready to run since the parameters
for individual transactions are not yet known. Hence, this is still a ``template'' --- here we call it the ``agreement template''
\item
For each new transaction under this agreement there will be a written or electronic Confirmation document.
In the simplest case this will do no more than provide value bindings for variables (the ``transaction parameters'') that are currently undefined
in the agreement template.
\item
A copy of the agreement template is made for each new transaction, and the transaction parameters for a transaction are passed 
as arguments to the code.  Additional parameters might also be passed, for example a unique identifier that
can be used to retrieve the original signed legal documents in the case of dispute.  The final version of the
smart contract code is then instantiated to run on a distributed ledger platform.
\end{itemize}

\section{Validating Smart Contract Code for Smart OTC Derivatives}

Many observers have pointed to the need for verification and validation of smart contract code\cite{khalil2017,harley2017,ISDALinklaters,Magazzeni}.  
In particular, \cite{Magazzeni} highlights five categories of verification and validation for smart contracts.  Here we focus on just one of those categories, which we find to be especially problematic --- Category 2 ``Does the computer program correctly encode the written natural language contract?''  

Validation may be substantially more difficult than verification. Whereas verification of smart contract code may aim to eliminate error states when the code is run,  validation of smart contract code aims to align the semantics of the legal documentation with the semantics of the code (e.g. does it correctly track the rights and obligations of parties, the discharging of obligations, the enforcement of prohibitions, the adherence to temporal aspects, and so on).  Understanding the semantics of the legal documentation is non-trivial, since it requires specialist knowledge in the two fields of banking and law; understanding the semantics of the code is similarly non-trivial, since it requires specialist knowledge in computer science.  The use of specialist terminology (including cases where common words may have specialist meanings) and implied (unspoken) knowledge has already led to misunderstandings between experts and effective validation may require investment into
the training of new staff as hybrid experts in the three areas of banking, law and computer science \cite{SCT-R3-3,ClackJDB}.

We have previously stated that a formal semantic specification of the legal documentation must include at least the temporal, deontic and operational aspects of that documentation. Yet that
formal specification must itself be validated to determine whether it correctly captures the meaning of the agreement.  This will require an expert in law (for example to disambiguate between temporal and non-temporal uses of the phrases such as ``will'', ``pursuant to'', 
``after giving effect to'' and ``after taking into account'',
or to distinguish between discrete and continuous periods of time as explained below), and also an expert in formal logic.  This validation of the specification will be facilitated if the formal specification is similar in structure to the legal text.
Three key issues arise:

\begin{enumerate}
\item
The {\em separability problem} --- the temporal, deontic and operational logics are closely intertwined and very difficult to separate, as explained in \cite{SCT-R3-3} and demonstrated further in Section~\ref{events}. 
\vspace{1mm}
\item
The {\em isomorphism problem} --- the structure of the semantic specification may be substantially different to the structure of the legal documentation, making it difficult for a specialist in law to understand and verify the semantics.\footnote{For example, a single legal clause may be represented by more than one expression in the formal semantics (perhaps distant from each other), and {\em vice versa} two or more legal clauses might be represented by a single expression in the formal semantics.}
\vspace{1mm}
\item
The {\em canonical form problem} --- there may be many different ways to structure the semantic specification for a given legal agreement; specifically there may be no unique normal form (``canonical form''), and this makes it difficult to compare two specifications to see if they are the same.\footnote{The existence of a unique canonical form (computable in reasonable time)
will depend on the logic employed, and the properties of its operators.  Where many logics are combined the existence of a unique canonical form may become problematic.}
\end{enumerate}



%
%


To achieve the aim of using formal methods to validate smart contract code for a Smart OTC Derivative transaction it will be necessary to create a formal represention of the whole agreement, including the Master Agreement, the Schedule and the Confirmation.  The Schedule and Confirmation may include freshly drafted provisions; thus, it will be necessary not only to represent the legal phrases that exist in the ISDA document set but also to anticipate provisions and constuctions that might appear in freshly drafted legal text.  

As a first step, we focus on the 2002 ISDA Master Agreement.  To simplify further, we start by exploring just the temporal aspects of the ISDA Master Agreement (though, due to the {\em separability problem}, we expect also to touch on deontic and operational aspects), resulting in the following structured observations.
These are then summarised in Section~\ref{requirements} as a set of requirements to guide the selection or design of a suitable temporal logic.

\subsection{Continuous and Discrete Time}
In the 2002 ISDA Master Agreement time is sometimes a continuous quantity, and sometimes discrete.

\vspace{-4mm}
\subsubsection{Continuous Time}
is a time interval typically used for prohibitions (e.g. a party is prohibited from doing something ``at any time''), normally expressed as a range with a start and an end date, and where such a range is not expressed it may be possible to infer a range from the textual context --- e.g.  the term of the Agreement.  Typical relevent phrases in the legal text are:
\begin{itemize}
\item
``with effect from'' may 
specify the start date of a continuous time range.
\item
``at all times until X'' specifies a continuous range with end date X.
\item
``so long as'' specifies a continuous range that persists for the duration of some other defined time span.\footnote{Which might be the duration of an obligation --- see Clause 4.}
\item
``to maintain in full force and effect all X'' refers to a continuous-time obligation (which might alternatively be modelled as a continuous-time prohibition against doing
anything that would negate any X).
\item
``in the future'' generally refers to a time period encompassing all times after the current (according to context, the end date might be the end of the agreement or transaction, or there may be no end date)
\item
``will survive'' generally indicates a continuous time period that continues after the end of the agreement or transaction (according to context).
\end{itemize}

\vspace{-5mm}
\subsubsection{Discrete Time}
is by contrast typically measured in days and can be expressed as  a single value, an ordered set of values, or a bag of alternative values:

\paragraph{A Single Discrete Time Value}
is a date representing a day (which might be before the effective date of the agreement, during the term of the agreement, or after the termination of the agreement).  Dates may be named, may be referenced via name or context (e.g. ``on that date'', ``on such date'', ``the date so designated'', ``the date specified'', ``the time specified'', ``the date determined under Clause X''), may be compared for time ordering (e.g. ``prior to'', ``the same day'', ``after'' and ``following''), may be counted (e.g. ``X days''), that count may be given a lower or upper bound (e.g. ``at least X days'', ``no more than Y days''), a date may be defined in relation to another date (e.g. ``at least 5 days after X''), a date may be subtracted from another date to give a number of days, and a number of days may be added to or subtracted from a date to give another date.  

A named date may have a value that is provided dynamically during performance of the agreement.  For example, the Early Termination Date may be ``designated'' during performance, and a date that is  ``otherwise agreed'' permits a date to be specifed by the parties in some other unspecified way which may be at the start of or during performance of a transaction. A date value can be tested to determine whether or not it was set in the legal text (e.g. using the phrases ``is specified in'' and ``is not specified in''), and whether it has been set during performance of the contract (``has been designated'').  Where a designated value replaces a previous value, it is necessary to retain previous values and the reason(s) for the designation(s) in order to support contractual phrases such as ``where an earlier Early Termination Date has been designated'',  ``Upon the occurrence or effective designation of'', and ``in the event of an Early Termination Date which is designated as the result of a Termination Event''. 
 
%

A discrete time value may have a number of associated properties.  For example, the property ``General Business Day'' refers to any day on which commercial banks are open for business. 
The phrase ``the first General Business Day after X'' therefore refers to the earliest date Y such that Y occurs after X and where Y has the associated property ``General Business Day''.\footnote{The property  ``General Business Day'' is described in a generic way so that it could potentially apply to an infinite number of dates, but a property could also be described in a way that
it could apply to only a finite number of dates.}
Another example property could be ``Designated Date'' or ``Designated Date Due to a Termination Event'' (since the text may state that a provision holds if a date has such a property).
The author is not aware of any specific discipline relating to the setting and testing of properties of time values within legal text, yet within a formal specification this would be an obvious area for
checking correctness (e.g. to ensure that if a property is checked it should be set at some other point in the agreement).

In the legal text, a specified date ``has occurred'' if the current date (during performance of the agreement) is the same day as or after the specified date.  By contrast, the legal phrase ``there is'' must be interpreted in context; sometimes it refers to the existence of a thing (which is not a temporal property) and at other times it may refer to the current time and may for example establish a reference date for a condition.

\paragraph{A Set of Discrete Time Values} 
is a time-ordered set of discrete dates with a start date and an end date, without duplicates: 

\begin{itemize}
\item
A set may be named and may be specified with certain days missing from the set (e.g. because they lack a given associated property).  
\item
A set may be defined in relation to a date (e.g. ``all days within 5 days after X'').
\item
A set of date values may be specified using a constrained universal quantifier (e.g. ``all days after event X and before event Y'').
\item
A set of specific date values may be specified using a formula representing repeated dates (e.g. ``every first Monday of every month'').
\item
A set of date values may start at the end date of an event (see Section~\ref{events}) and continue until Y days later (e.g. ``following event X, party A may terminate the agreement with no more than Y days notice'').
\item
``with effect from'' may (according to context) specify the start date of a discrete time period (i.e. of a set of discrete time values).
\end{itemize}

In the legal text, many phrases are used to introduce a set of discrete time values.  For example:
\begin{itemize}
\item
``notice requirement'', ``applicable grace period'' and ``applicable waiting period''  generally refer to a span of days (a set of discrete time values) where the start and end dates are normally expressly stated.
\item
``on any day'' may indicate a set of discrete dates (especially where it is followed by a qualification of the start or end dates, or both), following which each such day
may be referred to using the phrase ``on that day'' and certain days may be excluded from consideration using a phrase such as  ``(in each case, other than $\ldots$)''.
\item
``next succeeding Scheduled Settlement Date'' refers to a set of possible dates (Scheduled Settlement Dates), and selects that date which immediately follows the current date.
\item
``the time or times specified'' is a reference either to an individual date (which could be drawn from a representative bag of date values --- see below) or more commonly to a set of relevant specified dates.
\end{itemize}

\paragraph{A Bag of Alternative Discrete Time Values}
is a collection of alternative dates, which may contain duplicates, and which arises from phrases that permit a thing to occur on more than one date where two or more of those dates (perhaps specified relative to different events) may be the same.  For example, consider the phrase ``on or as soon as reasonably practicable following X'', which permits an action to occur on date ``X'' {\em or} on another date soon after ``X'' (see also Section~\ref{events} below for a discussion of reasonableness).

A bag of alternative discrete time values represents a single date, but the value of that date is generally only determined dynamically when the contract is performed.
This  introduces some complexity, since (i) prior to (or after) this date actually means prior to (or after) the date actually chosen rather than prior to (or after) the earliest (or latest) of the alternative dates, and (ii)  it is not yet clear whether it might be possible to constuct nested phrases ((prior to (X or Y)) OR (following (P or Q))) and so on.  This needs further attention.

\subsection{Temporal Aspects of Events, Obligations and Rights}
\label{events}

Events (an operational aspect of the agreement), obligations (a deontic aspect of the agreement) and rights, powers or privileges (also deontic aspects) all have associated temporal properties.  Thus, it is extremely difficult to separate temporal aspects from deontic and operational aspects (this is the aforementioned {\em separability problem}).

\vspace{-4mm}
\subsubsection{Events}
each have a start date and an end date (to support phrases such as ``has occurred'', ``is continuing'', and ``has ceased'', which can also appear in logical conjunction such as ``has occurred and is continuing''). The concept of an event can be used quite generally and may include for example (i) actions, such as the giving of notices; (ii) external events, such as the obtaining of judgement on an aspect of the agreement; and (iii) contract states, such as a party being in default, or a failure (e.g. to pay or to deliver).  Thus, events may be specified within the agreement, or may be created during the modelling process in order to construct the formal specification.


Any action that has an associated time could be an event, including passive actions such as becoming aware of a fact. For example the phrases ``upon becoming aware of'' and
``when the obligation is ascertained'' indicate that becoming aware and ascertaining are events with an associated time. The phrase ``the date of the information'' may according to context refer to the date of sending or receiving information, both of which are actions.\footnote{Conceivably this phrase might be used to refer to a date associated with a document, which raises the further issue of associating objects with temporal values.}

Defined events may or may not occur during the performance of the agreement.  Events also have other associated properties --- e.g. ``an event of default'' --- and there may be a total or partial ordering relationship applicable to events (though this is outside the scope of this paper).

In the legal text, a large number of phrases are used to link temporal properties with events.  For example:
\begin{itemize}
\item
``as of the time immediately preceding'' or ``immediately before'' an event X normally means the day before the start date of event X.
\item
``immediately'' normally means either the same day as, or the next day following, the occurrence of an event.  
\item
If an event X ``occurs prior to'' event Y this is generally taken to mean that the end date of X is prior to the start date of Y.
\item
``in such event'' generally refers to the immediately preceding named event.
\item
``the occurrence of'', ``the date as of'' and ``at such time of being'' each refers to an event and may according to context refer to the start date of an event or end date of an event or to the existence of a start date or end date for an event, and must be determined precisely from context.
\item
``at such time of being'' is the time at which an event occurs.
\item
``upon reasonable demand'' specifies the time of an event (a demand) with the proviso that the demand must be reasonable (which may have a temporal aspect such as occurring within or at a reasonable time).
\item
A ``potential event of X'' is an event which might become an event with property X (e.g. default) due to further events or the lapse of time.
\item
An event ``would occur'' as a result of some action if the action (e.g. performing an obligation, or entering into an agreement) would necessarily lead to the occurrence of the event.  This may be difficult to model, e.g. where the action or event is external to the agreement and in one possible future.\footnote{In the context of 
the ISDA Master Agreement it might sometimes be preferable to phrase this as a continuous prohibition to engage in an event that generates a Potential Event of Default or Potential Termination Event.} Similar problems occur with phrases such as ``would have been'' or ``would have been $\ldots$ if it were not for $\ldots$''.
\item 
``has taken action X'' could be represented by modelling that action as an event (thus it would have a start date and an end date, which are set dynamically during performance of the contract, and which may be the same).
If the end date of that event is before the current date (or other reference time according to context) then ``has taken action X'' would be true.
\item
``for so long as that is the case'' and ``for so long as the relevant event or circumstance continues to exist'' may both refer to an event, especially where a circumstance or the duration of a conditional can be modelled as an event (for example, given a phrase such as ``for so long as the party is unable to receive delivery'' would imply that being unable to receive deliery should be modelled as an event with start and end dates).
\end{itemize}

Applicable law and applicable
corporate policies might also be modelled as events in order to support phrases such as ``any applicable law $\ldots$ then in effect'', "party's policies in effect at that time" --- however, since there might be a very large number of such laws and policies, it might be better to replace such a provision with a call for human input to establish whether the stated condition (relating to law or policy) holds.

It is also important to note that sometimes the word ``event'' is not intended to refer to a thing that happens during performance of the contract, but rather to the specification of the contract itself and refers instead to the over-riding of one provision by another.  For example, the phrase ``in the event of any inconsistency'' is generally used to refer to an inconsistency between provisions of the agreement rather than a date during the performance of that agreement, and is followed by an indication of which provision should prevail over the other.



Less straightforward references to the temporal aspects of actions and events include the adverbs ``timely'' and ``promptly''.  These phrases rely on a court of law to 
apply a post-hoc test of reasonableness.   Similar phrases include ``as soon as is reasonably practicable'' and ``as soon as practicable''.  
For the purposes of validation, these adverbial phrases could be set to a global value such as ``within 1 day'' or ``within 2 Local Business Days''  (i.e. a set of discrete time values) during simulation of contract performance, to determine the extent to which the agreement might be sensitive to variations in such ambiguous time periods.

\subsubsection{Obligations}
\label{Obligations}
have a start date when the obligation is incurred, a due date, and a discharged date (if the discharged date is after the due date a sanction may be applied).
They also have (i) an optional end date at which point the obligation is automatically discharged if not previously discharged, and (ii)
an ordered set of zero or more ``revised due date(s)'' (used when an obligation has been deferred or accelerated). 
It is noted that the triggering of an obligation might also itself constitute an event.

The specification of repeated obligations may require an action to occur ``at least X times'', ``no more than Y times'' or ``at least X times but no more than Y times'' within a certain time interval.  Whilst the time interval itself can be expressed as a set, the specification of repeated occurrences is not really a temporal matter.  It is an example of how an obligation (deontic aspect) can be linked to repeated actions (operational aspect) within a defined time interval (temporal aspect).

In the legal text, example phrases that link temporal properties with obligations include:
\begin{itemize}
\item
``the due date'' and ``when due'' both refer to the due date of an obligation.
\item
``the last payment date'' or ``the last exchange date'' may according to context refer to either (i) the most recent such discharged date (e.g. the date of the most recently made payment), or (ii) from a set of due dates of such obligations to pay or exchange, the date that is latest.
\item
``satisfying a liability'' is generally a synonym for ``discharging an obligation''.
A party ``has satisfied'' some obligation (e.g. either to another party or to pay tax to an external body) if the current time is after the time that such obligation was discharged.  Note that obligations to external bodies may not be precisley expressed in the legal text and may need to be inferred from a provision that refers to the discharge of such an obligation.
\item
``will be deferred to, and will not be due until'' means the dynamic update of an obligation so that its revised due date is set to the stated value.
\end{itemize}

\vspace{-5mm}
\subsubsection{Rights, Powers and Privileges}
may apply throughout continuous time, or might become activated by the occurrence of a date or an event.  On the occurrence of such a date or event
it will be necessary to record that the associated right, power or privilege has been activated, and then also to record the date at which such right, power or privilege was exercised in relation to that date or event.  This supports phrases that refer to a delay in exercising a right, power or privilege.
Since a triggering event might occur many times, the activation time and exercise time should be recorded in each case.   
\vfill

\pagebreak
\section{Temporal Representation}
Hvitved \cite{hvitved12phd} provides a review of semantic techniques to support formal representation of legal agreements.
Our aim is to utilize a formal representation that can combine at least the deontic, temporal and operational aspects of standardised OTC derivatives contracts,
and of those technques surveyed by Hvitved the most attractive candidate is the technique developed by Lee \cite{lee}.  However Lee's representation of
temporal aspects needs to be expanded to cover the complexity demonstrated above.  Here we briefly discuss Lee's temporal framework and then set out
an intial set of requirements for extending the framework.

\subsection{Temporal logic in Lee's framework}

The Rescher and Urquhart temporal logic system \cite{rescher} is the basis for representing temporal aspects in Lee's framework.  
Lee calls this the ``RU calculus'' and explores and extends the use of this system. Pithadia \cite{pithadia} provides an initial critical assessment of Lee's framework.

\subsubsection{The Rescher and Urquhart system} is based on the temporal operator $R_t\Phi$ that denotes $\Phi$ (a formula in conventional first order logic with identity)
being realized at time $t$.  Lee rehearses the 
axioms for $R$, the total ordering relational operator $U$, a function $f :: Time \rightarrow {\rm I\!R}$, 
and temporal addition $\oplus$, as follows:
\[
R_t (\neg \Phi) \leftrightarrow \neg R_t\Phi  
\]
\[
R_t(\Phi \& \Psi) \leftrightarrow R_t \Phi \& R_t \Psi
\]
\[
R_{t'}(\forall t \Phi) \leftrightarrow \forall t R_{t'} \Phi
\]
\[
U_{t t'}        \;\; \textit{indicates that time t precedes time t'}
\]
\[
f(t \oplus t') = f(t) + f(t')
\]

Lee modifies the RU calculus by introducing the day as the basic unit of time and by introducing the concept of time intervals 
where $SPAN(d, d')$ defines a time span from the beginning of date $d$ to the end of date $d'$, $BEG(d)$ gives the beginning date of a time interval, 
and $END(d)$ gives the end date of a time interval.  Two operations are introduced on time intervals (using Lee's notation, where $d$ is a time interval):

\[
RD_d \Phi \leftrightarrow \exists t (t \in d) \& R_t \Phi
\]
\[
RT_d \Phi \leftrightarrow \forall t(t \in d) \Longrightarrow R_t \Phi
\]

Thus 
$RD_d\Phi $ indicates that $\Phi$ is realized at least once during the time interval $d$ and
$RT_d\Phi$ indicates that $\Phi$ is realized throughout time interval $d$.  A further operator $RB_D \Phi$ indicates that $\Phi$ is realized before day $D$
and is defined in terms of the $RD$ operator using an arbitrary undefined day in the past, which Lee denotes as ``$\_$'' and we interpret as a ``bottom'' element  ``$\bot$'':
\[
RB_D \Phi = RD_{SPAN(\bot,D)} \Phi
\]

Lee further observes that a calendar of dates is an interval scale with
no obvious value of ``zero'' and therefore addition is more complex than represented in the RU calculus; he redefines the operator
$\oplus$ to take a date $d$ and a number of days $n$ and return the date that is $n$ days later than $d$, as follows (where $D$ is the unit ``days''):
\[
d \oplus nD
\]

\subsubsection{Lee's temporal framework} has a number of limitations in the context of smart OTC derivatives, 
In overview, the most obvious shortcomings are (i) the inability to specify or reason about continuous time; (ii) the inability to specify dates with different properties (such as ``Business Day''); and
(iii) the inability to specify concisely a set of times with subtle membership rules (such as ``the first Friday of every month'').   Lee's time intervals provide an implementation of sets (and possibly bags, depending on context) but it is very difficult to specify a time interval with specific dates missing.
Whilst it is possible to use a combination of RU calculus operators and Lee's operators to represent phrases such as ``action X occurs Y days after event Z'' (which could for example be represented as 
$R_t Z \& (R_{t \oplus Y} X)$), it is difficult to express a provision such as ``payments made at a weekend will be processed on the first Business Day of the following week''.

Lee's temporal framework could be modified and extended, or could be replaced.  However, it is not sufficient to make such a decision based only on an analysis of the temporal aspects.  
The great advantage of Lee's framework is the way that it combines deontic, operational and temporal aspects.  We therefore eschew construction of a formal temporal logic until 
analysis of the deontic and operational aspects of the ISDA Master Agreement has been conducted.  Instead, we provide an initial outline of requirements in the following section.

\subsection{Initial Requirements for the Temporal Aspects of a Semantic Framework}
\label{requirements}
Given the {\em separability problem} relating to the very close coupling between temporal, deontic and operational aspects of legal documents, it is unlikely that a separate temporal logic would be appropriate for formal modelling of smart OTC derivatives.  In most cases time is a {\em property} of deontic and operational aspects (the current time is a notable exception to this observation).  However, we can summarise our investigation of the ISDA Master Agreement by setting out some guideline requirements for the expressibility of the temporal parts of the combined logic, as follows:

\subsubsection{Requirements for Continuous Time Intervals.}

It must be possible to express intervals of continuous time with a start and end point denoted by a discrete time value.  In general, we would wish to support the following operations:
\begin{itemize}
\item
create a new continuous time interval with discrete time values for the start and end points (if the end-points are defined to be {\em outside} the interval, this would give a straightforward representation of an ``empty'' interval as being one where the start and end points are the same, and it would not be an error to request the start or end point of an empty interval);
\item
bind a name to an interval;
\item
create an aggregate collection of intervals (this is one way to implement a union of non-overlapping intervals, since we cannot have an interval result containing gaps);
\item
get the start or end point of an interval;
\item
get the intersection of two intervals (perhaps returning an ``empty'' interval);
\item
test whether a discrete time value is before the start of such an interval; and 
\item
test whether a discrete time value is after the end of such and interval.  
\end{itemize}

Although in some legal prose a period of continuous time might have a start point that is infinitely in the past (which we denote $T_{-\infty}$), or an end point that is infinitely in the future (which we denote $T_{\infty}$), it is not yet clear whether this is essential or merely lazy drafting.  It is an open question as to whether the temporal logic should support such ``extreme'' values for start and end points of continuous time intervals (nor whether they might be useful for other purposes).  

\subsubsection{Requirements for Single Discrete Time Values.}

It must be possible to express single discrete time values, to associate {\em properties} with a single discrete time value, to keep a history of updates to the time value that is bound to a name, to calculate
differences between two dates as a number of days, and to create and use date expressions (including the use of days as relative offset values).

In general, we would wish to support the following operations:

\begin{itemize}
\item
create a new single discrete time value;
\item
provide the special single discrete time value that is the {\em current time};
\item
bind a name to a historical list of single discrete time values (and with each such value record when it was bound, who by, why, whether it was bound in the text or during performance, and perhaps some further properties);
\item
perhaps provide the extreme values $T_{-\infty}$ and $T_{\infty}$ mentioned above and to test whether a single discrete time value is one of these two extreme values;
\item
increment a discrete time value by one day;
\item
decrement a discrete time value by one day;
\item
get the difference in days between two discrete time values (though this deserves more attention, since we may need to calculate the number of days with a specified property - e.g. Business Days);
\item 
associate a set of properties with a discrete time value and test a discrete time value to see if it has a stated property;
\item
apply a predicate to a discrete time value to see if it passes or fails a test;
\item
provide equality and relational operators to use on two dates;
\item
create and use date expressions combining any of the above operations.
\end{itemize}

\subsubsection{Requirements for Sets or Bags of Discrete Time Values.}

It must be possible to express a set or bag of discrete time values where a set or bag has a start and end date and a (possibly discontinuous) collection of valid dates between the start and end dates.  
It must also be possible to define the members of the set or bag using a generator expression.  As with continuous time intervals, if the start and end dates are defined to be {\em outside} the set or bag this would give a straightforward representation of an ``empty'' set or bag as being one where the start and end points are the same, and it would not be an error to request the start or end point of an empty set or bag.

In general, we would wish to support the following operations:

\begin{itemize}
\item
create a new set or bag of discrete time values;
\item
get the start date or end date of a set or bag;
\item
test whether a given date is a member of the set or bag;
\item
get the intersection of two sets or bags, returning a set or bag respectively (which might be empty);
\item
get the union of two sets or bags, returning a set or bag respectively;
\item
test whether two sets or two bags are equal;
\item
bind a name to a set or bag;
\item
apply a filter to a set or bag, to produce a set or bag (respectively) that may be smaller or equal in size;
\end{itemize}

\subsection{Temporal Requirements Relating to Deontic and Operational Aspects.}

Each event must have a start date and an end date.  In the legal text, it is possible that these dates might be specified in relation to some other dates or as a set or bag of possible dates.  However, once the event has started any previous ``possible'' start time must be updated with the actual start time (and similarly for the end time).  Thus when using a temporal logic operator such as $R_t \textit{E}$ (where $\textit{E}$ is an event) it is necessary to disambiguate between the ``realisation'' of the possible and actual start of the event $\textit{E}$ and the ``realisation'' of the possible and actual end of that event.

As explained in Section~\ref{Obligations}, each obligation has a start date, an optional end date, a due date, and a discharged date.  Thus when using a temporal logic operator such as $R_t \Phi$ (where $\Phi$ is an obligation) it is necessary to disambiguate between the realisation of the incurring of the obligation, the passing of the end date, the passing of the due date without discharge, or the realisation of the discharge of the obligation.  Similarly, rights, powers and privileges have activation times and (perhaps multiple) exercise times and these must be disambiguated.

Where deontic or operational aspects may be repeated a set, minimum, or maximum number of times, there must be a mechanism to express these repetitions.

\section{Summary and Conclusion}

The legal language used to draft standardised contracts for OTC derivatives is rich and complex.  Yet it is essential to have a full and formal understanding of the semantics of the legal documentation
in order to validate the associated smart contract code.  Here we have presented initial results from an analysis of the temporal language used in the 2002 ISDA Master Agreement for OTC derivatives.

When a formal semantic description of the legal documentation is constructed, it must specify not only the temporal semantics but also the semantics of (at least) the deontic and operational aspects.
The {\em separability problem} --- the fact that the temporal, deontic and operational logics are closely interwined --- requires a combined logical framework.  The results of an analysis of the temporal language used in legal documentation are therefore presented as a set of initial requirements.  More work is required to analyse the deontic and operational language, each of which will lead to its own 
set of requirements.  Our final aim is to combine these into a single semantic framework.  This is a formidable challenge, yet we envisage that such a framework will be applicable not only to the automation of OTC derivatives but also to the formal understanding of a wide range of legal contracts and statutes.

\subsection{Acknowledgements.}
The authors are grateful to UCL students Justin Jude and Mengyang Wu who assisted this work by reviewing logic frameworks and providing supporting tools.
\vfill

\bibliographystyle{splncs04}
\bibliography{Clack-iSOLA-2018}
%
%
%
%
%
\end{document}